\documentclass[aps,prb,twocolumn,superscriptaddress,amsmath,amssymb,longbibliography]{revtex4-1}
\usepackage{newtxtext,newtxmath}
\usepackage{graphicx}
\usepackage{bm}
\usepackage[colorlinks=true,citecolor=blue,breaklinks=true,linkcolor=red,
linktocpage=true,urlcolor=blue,pagebackref=false]{hyperref}
\newcommand{\ev}[1]{\ensuremath{\langle #1 \rangle}}
\newcommand{\abs}[1]{\ensuremath{|#1|}}

\newcommand{\ket}[1]{\ensuremath{|#1\rangle}}

\newcommand{\tr}{\ensuremath{\operatorname{tr}}}
\newcommand{\sgn}{\ensuremath{\operatorname{sgn}}}
\newcommand{\diag}{\ensuremath{\operatorname{diag}}}
\newcommand{\up}{{\uparrow}}
\newcommand{\down}{{\downarrow}}
\newcommand{\id}{{\ensuremath{\openone}}}
\renewcommand{\Im}{\ensuremath{\operatorname{Im}}}

\newcommand{\void}[1]{}

\begin{document}
	\title{Resonator induced quantum phase transitions in a hybrid Josephson junction}
	
	\author{Robert Hussein}
	\affiliation{Institut f\"ur Festk\"orpertheorie und -optik, Friedrich-Schiller-Universit\"at Jena, D-07743 Jena, Germany}
	\author{Wolfgang Belzig}
	\affiliation{Fachbereich Physik, Universit\"at Konstanz, D-78457 Konstanz, Germany}
	\date{\today}
	
	\begin{abstract}
		We investigate the Josephson current through a suspended carbon nanotube double quantum dot which, at sufficiently low 
		temperatures, is characterized by the ground state of the electronic subsystem. Depending on parameters like a magnetic 
		field or the inter-dot coupling, the ground state can either be a current-carrying singlet or doublet, or a blockaded triplet state.
		Since the electron-vibration interaction has been demonstrated to be electrostatically tuneable, we study in particular its effect 
		on the current-phase relation. We show that the coupling to the vibration mode can lift the current-suppressing triplet blockade 
		by inducing a quantum phase transition to a ground state of a different total spin. Our key finding is the development of a triple 
		point in the Josephson current parameterized by the resonator coupling and the Josephson phase. The quantum phase transitions 
		around the triple point are directly accessible through the critical current and resilient to moderately finite temperatures. The 
		proposed setup makes the mechanical degree of freedom part of a superconducting hybrid device which is interesting 
		for ultra-sensitive displacement detectors.
	\end{abstract}
	
	\maketitle
	
	\section{Introduction}
	The tunnel coupling of quantum dots (QDs) to superconducting leads can open Andreev bound-state channels that may carry a 
	supercurrent~\cite{PilletNatPhys2010a,LeeNatNano2014a,HwangPRB2015a,HwangEPJB2017a} and, thus, realize Josephson-like 
	junctions~\cite{ChoiPRB2000a,KarraschPRB2008a,MengPRB2009a,DeFranceschiNatNano2010a,Martin-RoderoAP2011a,WentzellPRB2016a} 
	that might even realize a nontrivial topology. \cite{RiwarNC2016a,KleesPRL2020a,MeyerArXiv2019a,WeisbrichPRXQuantum2021a}
	The QDs' highly tunable electronic spectra~\cite{CiorgaPRB2000a,Tarucha2001a,WielRMP2002a,BiercukNanoL2005a} makes them in 
	particular interesting for low-dissipation superconducting spintronics~\cite{EschrigRPP2015a,LinderNatPhys2015a} and quantum 
	computation.~\cite{LaddNature2010a,AruteNature2019a} Ground-state (GS) transitions in such systems have been 
	observed~\cite{DamNature2006a,DelagrangePRB2016a,DelagrangePhysicaB2018a,EstradaSaldanaPRL2018a} and discussed in terms of 
	the spin-orbit interaction~\cite{LimPRL2011a,DrosteJP2012a}, topological 
	protection,~\cite{MarraPRB2016a,TiiraNatureC2017a,MarraBeilsteinJN2018a,ArracheaPRB2019a,BlasiPRB2019a} and nonequilibrium 
	transport.~\cite{PalaNJP2007a,GovernalePRB2008a,FuttererPRB2013a,OguriPRB2013a} Remarkably, it was recently demonstrated 
	in carbon nanotube (CNT) quantum dot setups that the electron-phonon interaction can be tailored electrostatically, to even 
	attain attractive electron interaction.~\cite{KuemmethNature2008a,BenyaminiNP2014a,HamoNature2016a} In combination with 
	superconducting leads, this may enable the design of novel quantum states of matter.~\cite{MarganskaPRL2019a,BlienNC2020a}
	\begin{figure}[b!]
		\includegraphics[width=\columnwidth]{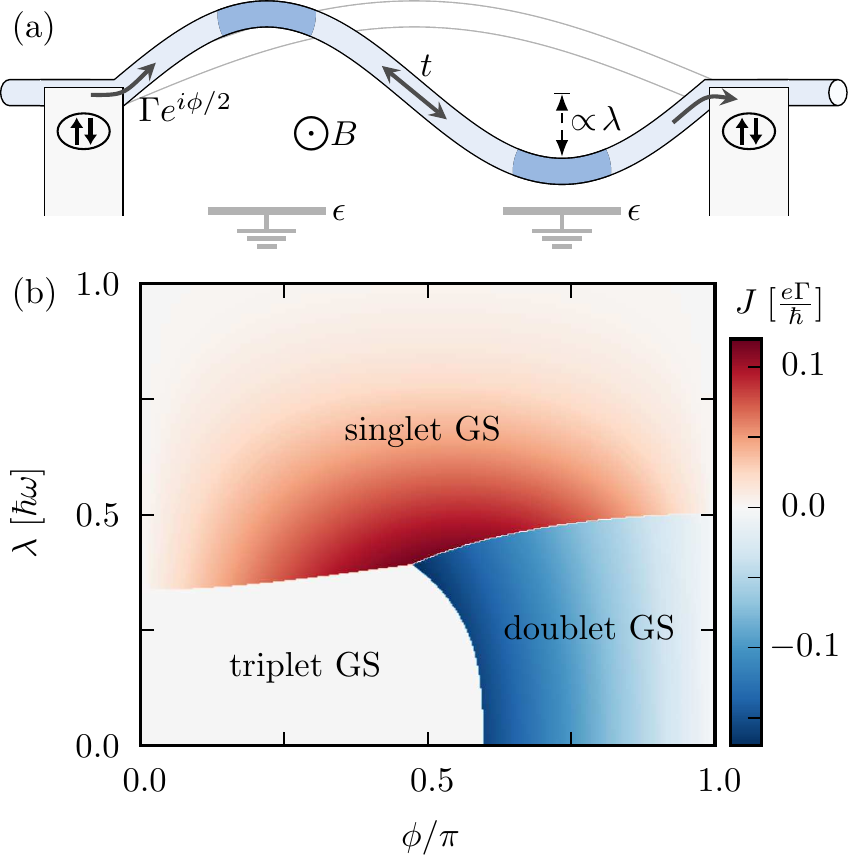}\caption{\label{fig.:scheme}(a) Carbon nanotube quantum dot subject to a Zeeman field $B$ and coupled with the rate $\Gamma$ to two 
		superconducting leads with phase difference $\phi$. Depending on the system's ground state, Cooper-pairs may tunnel from one superconductor, 
		via the double quantum dot, to the other establishing, thus, a Josephson current. The coupling $\lambda$ to the resonator renormalizes the onsite 
		energies $\epsilon$ of the dots and may change the total spin $s$ of the system.
		(b) Josephson current $J$ in the ($\phi$,$\lambda$)-plane at zero temperature, $k_B T=0$, for the intradot Coulomb interaction 
		$U_C=t=\Gamma=\hbar\omega$ and $B=0.9\:\Gamma$, evaluated at the particle-hole symmetric point,
		$\epsilon=-U_C/2$. It features quantum phase transitions between ground states of different total spin and a common triple point. 
		}
	\end{figure}
	CNT devices~\cite{SchindelePRL2012a,JungNanoL2013a,WenNP2019a,UrgellNP2019a,LotfizadehPRL2019a,KarkiPRB2019a} are characterized by their small masses and qualify, likewise, as conductors and nanoresonators 
	with high quality factors.~\cite{HuettelNanoL2009a,GanzhornPRL2012a,MoserNatureN2014a} They have potential applications 
	in lasing~\cite{LiOExpress2012a, LiuOMExpress2015a, MantovaniPRB2019a,RastelliPRB2019a} and serve in the ultra-sensitive detection of masses,~\cite{NaikNatureN2009a,ChasteNatureN2012a}
	charge densities,~\cite{ShapirScience2019a}
	magnetic moments,~\cite{CleuziouNatureN2006a,UrdampilletaNatureM2011a,PeiNatureN2012a} and terahertz frequencies.~\cite{RinzanNanoL2012a} In order to achieve noise reduction and single-phonon control, ground-state cooling of nanoresonators has been studied experimentally~\cite{OConnellNature2010a,ClarkNature2017a} 
	and theoretically.~\cite{StadlerPRL2014a,StadlerPRL2016a,MantovaniPRResearch2019a}
	
	In this Letter, we consider a suspended double quantum dot (DQD) resonator that is attached to two s-wave superconductors and exposed to a 
	magnetic field, see Fig.~\ref{fig.:scheme}(a). The latter controls the Zeeman splitting of the QD levels and results, for a sufficiently strong field, in
	a triplet ground state suppressing the Josephson current due to its incompatible spin-symmetry with the superconducting condensates.~\cite{EldridgePRB2010a}
	As shown below, we find that the coupling to the resonator can change the total spin of the QD system, and, thus, help to overcome 
	such a triplet blockade. The total spin quantum number serves as an order parameter and we show that it can give rise to a 
	triple point in the quantum phase diagram spanned by the Josephson phase and the resonator coupling. In particular, we illustrate in 
	Fig.~\ref{fig.:scheme}(b) that the corresponding quantum phase transitions between ground states of different total spin are directly reflected 
	in the Josephson current. The occurring triple point further manifests in the critical supercurrent, as investigated in Fig.~\ref{fig.:JCritical}. 
	To underline the quantum nature~\cite{VojtaRPP2003a, PeanoNewJP2014a} of the zero temperature phase transitions, we show in Fig.~\ref{fig.:finiteT} 
	that they smooth out, when thermal fluctuations start to play a role. The central transport features, however, are preserved for experimentally relevant temperatures.

	\section{Quantum dot Josephson junction}%
	\begin{table}[t] 
		\begin{ruledtabular}
			\begin{tabular}{lll}
				$\ket{0}$ & & empty state \\
				$\ket{dd}$ & $=d_{R\downarrow}^\dag  d_{R\uparrow}^\dag d_{L\downarrow}^\dag  d_{L\uparrow}^\dag \ket{0}$
				& fully occupied\\
				$\ket{S0}$ & $=\frac{1}{\sqrt{2}} \big( 
				d^\dag_{R\uparrow} d^\dag_{L\downarrow} 
				-d^\dag_{R\downarrow} d^\dag_{L\uparrow}
				\big)\ket{0}$ & nonlocal singlet\\
				$\ket{S\pm}\!$ & $=\frac{1}{\sqrt{2}} \big( 
				d^\dag_{L\downarrow} d^\dag_{L\uparrow} 
				\pm d^\dag_{R\downarrow} d^\dag_{R\uparrow}
				\big)\ket{0}$ & local singlet\\
				\hline
				$\ket{T0}$ & $=\frac{1}{\sqrt{2}}\big( 
				d^\dag_{R\uparrow} d^\dag_{L\downarrow}
				+d^\dag_{R\downarrow} d^\dag_{L\uparrow}
				\big)\ket{0}$ & mixed-spin triplet\\
				$\ket{T\sigma}$ & $=d^\dag_{R\sigma} d^\dag_{L\sigma}\ket{0}$ & equal-spin triplet\\
				\hline
				$\ket{\sigma\pm}$ & $=\frac{1}{\sqrt{2}}\big( 
				d^\dag_{L\sigma}\pm d^\dag_{R\sigma}
				\big)\ket{0}$& singly occupied\\
				$\ket{t\sigma\pm}\!$ & $=\frac{1}{\sqrt{2}}\big( 
				d^\dag_{L\sigma}d^\dag_{R\downarrow}d^\dag_{R\uparrow}
				\pm d^\dag_{R\sigma}d^\dag_{L\downarrow}d^\dag_{L\uparrow}
				\big)\ket{0}$& triply occupied
			\end{tabular}
		\end{ruledtabular}
		\caption{\label{tab.:basis}%
		Basis of the system Hamiltonian, subdivided into the singlet (top cell), triplet (middle cell), and doublet sector (bottom cell). 
		Here, $\sigma=\up,\down$ labels the electron spin.
		}
	\end{table}
	We study the DQD Josephson junction in the limit of a large superconducting gap and consider an Anderson-Holstein type~\cite{HolsteinAP1959a,AndersonPR1961a,HusseinPRB2010a,MetelmannPRB2011a}  system Hamiltonian,
	\begin{equation}
		H_{\textrm{S}}= H_{\textrm{DQD}} +\hbar\omega a^\dag a 
		+\lambda\big(a^\dag +a\big)\sum_{\sigma}\big(n_{L\sigma}-n_{R\sigma}\big), \label{eq.:HS}
	\end{equation}
	where the occupation imbalance, $n_{L\sigma}-n_{R\sigma}$, of the double quantum dot is coupled with the strength $\lambda$ to a 
	mechanical mode of frequency $\omega$. Here, $a^\dag$ is the creation operator of the bosonic mode, and 
	$n_{\alpha\sigma}=d^\dagger_{\alpha\sigma}d_{\alpha\sigma}$ denotes the fermionic occupation operator, where $d^\dagger_{\alpha\sigma}$ 
	creates an electron on the dot  $\alpha=L,R$ with spin $\sigma = \uparrow,\downarrow$.
	The double quantum dot is modeled by~\cite{RozhkovPRB2000a,DrosteJP2012a,HusseinPRB2016a,HusseinPRB2019a}
	\begin{equation}
		\begin{split}
			H_{\textrm{DQD}} {=}
			{ }& \sum_{\alpha\sigma}\Big(\epsilon+\sigma\frac{B}{2} \Big)n_{\alpha\sigma} 
			+\frac{t}{2} \sum_{\sigma} \big(d^\dag_{L\sigma}d_{R\sigma} +\mathrm{H.c.}\big)\\
			+&\sum_{\alpha}\Big[
			U_{C}\:\! n_{\alpha\up}n_{\alpha\down} - \frac{\Gamma}{2} \big( 
			e^{\alpha i\phi/2} d^\dagger_{\alpha\up} d^\dagger_{\alpha\down}+\mathrm{H.c.}
			\big)\Big]\label{eq.:Heff},
		\end{split}
	\end{equation}
	where the first term describes the Zeeman splitting  of the dot levels with energy $\epsilon$ due to an applied magnetic field $B$. 
	The second term characterizes the interdot tunneling with tunneling amplitude $t$. Each dot $\alpha$ can house up to two electrons 
	of opposite spin which are subject to the intradot Coulomb interaction $U_C$. The term $\propto d^\dagger_{\alpha\up} d^\dagger_{\alpha\down}$ describes the Andreev tunneling 
	of a Cooper-pair with the rate $\Gamma$ into dot $\alpha$, where $\phi$ is the Josephson phase. We use
	the convention $\sigma=\pm$ for the spin up/down and $\alpha=\pm$ for the left/right dot.
	The Josephson current~\cite{JosephsonAP1965a,BlochPRB1970a,Beenakker1992a,GolubovRMP2004a}
	\begin{align}
	J=\frac{2e}{\hbar}\partial_\phi F(\phi) \label{eq.:J}
	\end{align}
	through the system, is given by the derivative of the free energy  $F=-k_BT\ln Z$ with respect to the phase $\phi$. Here, $Z=\tr e^{-H_{\textrm{S}}/k_BT}$
	denotes the partition function of the canonical ensemble. Since the Josephson current is $2\pi$-periodic in $\phi$ and antisymmetric in $\phi\to-\phi$, 
	we restrict the Josephson phase  hereafter to the regime $0\leq\phi<\pi$. 
	The Josephson current is furthermore particle-hole symmetric, $\epsilon\to2\epsilon_0-\epsilon$, with $\epsilon_0\equiv-U_C/2$. 
	At zero temperature, $T=0$, Eq.~\eqref{eq.:J} reduces to $J=(2e/\hbar)\partial_\phi E_{\textrm{GS}}$ with
	$E_{\textrm{GS}}$ being the ground state energy of the system. The diagonalization of the system Hamiltonian $H_S$ and, therewith, the calculation 
	of the Josephson current is, however, hindered by the coupling to the oscillator. In the following, we will eliminate the oscillator degrees of freedom 
	and derive an effective low-dimensional Hamiltonian for the electronic subsystem.

	\section{Elimination of the oscillator}
	In order to eliminate the electron-phonon coupling term in the system Hamiltonian, Eq.~\eqref{eq.:HS}, and eventually the oscillator mode,
	we introduce the polaron transformation~\cite{LangJETP1962a,BrandesPRL1999a,Mahan2000a,BrandesPR2005a} 
	$\bar O \equiv e^X O e^{-X}$ with $X=\frac{\lambda}{\hbar\omega}\big(a^\dag-a\big)\sum_{\sigma}\big(n_{L\sigma}-n_{R\sigma}\big)$. From the transformations $\bar a = a-(\lambda/\hbar\omega)\sum_{\alpha\sigma}\alpha n_{\alpha\sigma}$ 
	and  ${\bar d}_{\alpha\sigma}=D\big(\alpha\lambda/\hbar\omega\big) d_{\alpha\sigma}$ of
	the bosonic and fermionic operators, respectively, one finds straightforward the transformed system
	Hamiltonian ${\bar H}_S$. Here, $D(z)\equiv\exp(z a^\dag-z^* a)$ denotes the displacement operator generating
	a coherent state when applied on the phonon vacuum. Finally, we assume a strong coupling to a thermal bath and take the average with respect to the
	thermal state $\rho_{\rm osc}\propto\exp\big(-\hbar\omega a^\dag a/k_BT\big)$ of the oscillator. Then, we obtain the effective Hamiltonian
	\begin{equation}
		H_{\textrm{POL}}  = \ev{{\bar H}_S}_{\rm osc}
		-\frac{\lambda^2}{\hbar\omega}\sum_{\alpha\alpha'\sigma\sigma'}\alpha\alpha' n_{\alpha\sigma}n_{\alpha'\sigma'} + \hbar\omega n_B(\omega).
	\end{equation} 
	The Bose function $n_B(\omega)=\big[\exp\big(\hbar\omega/k_BT\big)-1\big]^{-1}$ in the last term stems from the
	thermal average  $\ev{a^\dag a}_{\rm osc}$ of the phonon number operator. It vanishes in the limit of zero temperature. With the aid of the 
	identity $n_{\alpha\sigma}^2=n_{\alpha\sigma}$  and the relation $D(x)D(y)=D(x+y)\exp[i\Im(x y^*)]$, the effective Hamiltonian becomes
	\begin{equation}
		\begin{split}
			H_{\textrm{POL}} {=}
			{ }& \sum_{\alpha\sigma}\Big(\bar\epsilon+\sigma\frac{B}{2} \Big)n_{\alpha\sigma} 
			+\frac{t}{2} \sum_{\sigma} \big(d^\dag_{L\sigma}d_{R\sigma} +\mathrm{H.c.}\big)\\
			+&\sum_{\alpha}\Big[
			{\bar U}_C\:\! n_{\alpha\up}n_{\alpha\down} - \frac{\bar\Gamma}{2} \big( 
			e^{\alpha i\phi/2} d^\dagger_{\alpha\up} d^\dagger_{\alpha\down}+\mathrm{H.c.}
			\big)\Big]\\
			+&\bar U\sum_{\sigma\sigma'}n_{L\sigma}n_{R\sigma'} 
			+ \hbar\omega  n_B(\omega)\label{eq.:HPol}
		\end{split}%
	\end{equation}%
	with the renormalized onsite energies $\bar\epsilon=\epsilon - \lambda^2/\hbar\omega$,
	intradot Coulomb interaction ${\bar U}_C=U_C-2 \lambda^2/\hbar\omega$,
	and tunneling rate $\bar\Gamma  =\Gamma \ev{D(2 \lambda/\hbar\omega)}$. 
	The thermal average of the displacement operator is given by~\cite{DominguezPRB2011a}
	$\ev{D(z)} =\exp\big[-\big(\abs{z}^2/2\big)\coth\big(\hbar\omega/2 k_BT\big)\big]$.
	Notice the  emergence of the next-to-last term in Eq.~\eqref{eq.:HPol} describing an effective interdot Coulomb interaction 
	with coupling strength $\bar U =2\lambda^2/\hbar\omega$. If not stated otherwise, we assume the magnetic field to be positive, $B>0$. 
	Its application in the opposite direction, $B\to-B$, would just interchange the spin-up with the spin-down states, but leaves the Josephson current unchanged. 
	The effective Hamiltonian, Eq.~\eqref{eq.:HPol}, is block-diagonal
	in the singlet, the triplet, and the doublet sector listed in table~\ref{tab.:basis}, which differ in their total spin quantum number $s$. So in order to find the ground-state energy $E_{\textrm{GS}}$ of the system, one can just diagonalize each sector separately and then
	determine the lowest eigenvalue. At the particle-hole symmetric point $\epsilon=\epsilon_0$ one finds that 
	$E_T=2\epsilon_0-\abs{B}$ is the lowest eigenvalue of the triplet sector, and that $E_D=\frac{1}{2}\big(E_T-\bar U -\sqrt{\bar\Gamma^2+t^2+2\bar\Gamma t \abs{\sin(\phi/2)}}\big)$ is the lowest eigenvalue of the doublet sector. The lowest eigenvalue of the singlet sector can be estimated by
	\begin{align}
	E_S &\approx E_S^0 + \frac{
		\bar\Gamma^2\big(1+\cos\phi\big)\big(E_S^0-2\epsilon_0\big)
	}{
		4E_S^0 \big(E_S^0 -\epsilon_0+ \bar U \big)
	} \label{eq.:ESApproxMainText}
	\end{align}
	with $E_{S}^0 =\epsilon_0-\bar U-\sqrt{(\epsilon_0+\bar U)^2+t^2}$, see the  Appendix.
	The derivative in $\phi$ of the ground-state energy $E_{\textrm{GS}} = \min\{E_S,E_D, E_T\}$ yields, eventually,
	the estimate
	\begin{equation}
		J_{\textrm{est}}{\equiv} -\frac{e}{\hbar}\begin{cases}
			\dfrac{
			\bar\Gamma^2 \sin(\phi)\big(E_S^0-2\epsilon_0\big)
			}{
			2E_S^0 \big(E_S^0 -\epsilon_0 + \bar U\big)
			} & E_S < E_D, E_T\\
			\dfrac{\bar\Gamma t\cos(\phi/2)\sgn\bm{(}\sin(\phi/2)\bm{)}}{2\sqrt{
			\bar\Gamma^2+t^2+ 2\bar\Gamma t\abs{\sin(\phi/2)}
			}} & E_D < E_S, E_T\\
			0 & E_T < E_S, E_D
		\end{cases}\label{eq.:JEst}
	\end{equation}
	of the Josephson current at zero temperature. 
	
	\begin{figure}[t]
		\includegraphics{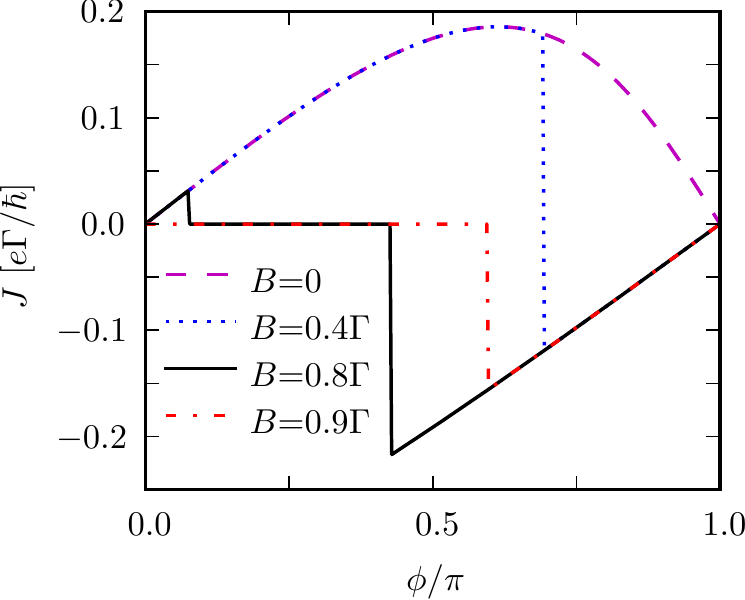}\caption{\label{fig.:phaseTransition}%
		Josephson current $J$ as a function of the phase $\phi$ for different values of the magnetic field $B$ and vanishing coupling to the oscillator, $\lambda=0$. 
		Parameters are $\epsilon=-U_C/2$,  $t=U_C=\Gamma$, and $k_B T=0$.}%
	\end{figure}%
	
	\begin{figure}[t]
		\includegraphics{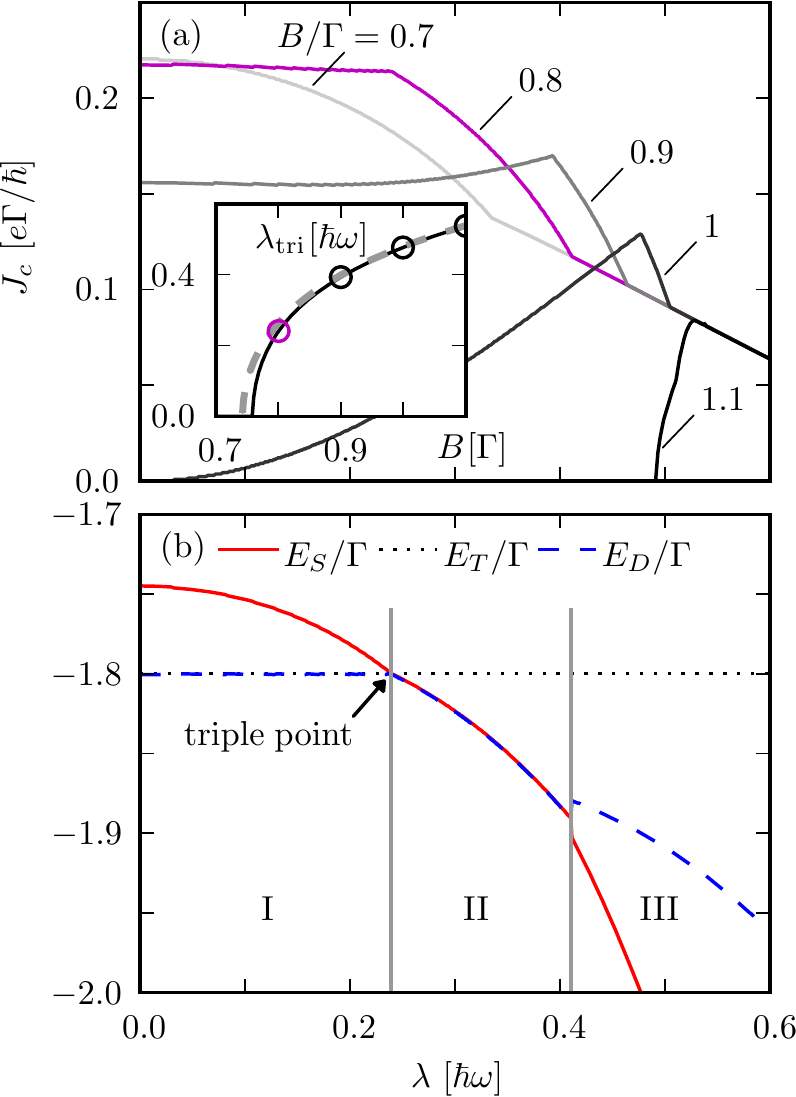}\caption{\label{fig.:JCritical}%
		(a) Critical current $J_c$ as a function of the oscillator coupling $\lambda$ for different values of the magnetic field $B$ and all other
		parameters as in Fig.~\ref{fig.:phaseTransition}. The inset shows the coupling strength $\lambda_{\textrm{tri}}$ as a function of the magnetic field
		at which all three ground states coexist. The dashed gray line corresponds to the estimate of the Josephson current given in Eq.~\eqref{eq.:JEst}
		and the circles correspond to the magnetic fields in panel (a) for which a triple point emerges.
		(b) Lowest eigenvalues of the singlet, triplet, and doublet sector corresponding to $B=0.8\:\Gamma$. The vertical lines indicate phase transitions between distinct ground states.
		}
	\end{figure}
	 
	\section{Induced quantum phase transitions and triple point}
	First, let us consider the case without the coupling to the resonator, $\lambda=0$, and restrict to zero temperature, $k_BT=0$. 
	In figure~\ref{fig.:phaseTransition}, we show the Josephson current $J$ in dependence of the phase $\phi$ and choose
	the parameters such that the QD system is in absence of a magnetic field in a singlet ground state (dashed line). As expected,
	$J$ is positive in the interval $0<\phi<\pi$, but shows deviations from the pure sinusoidal behavior due to  a finite
	occupation of the nonlocal singlet state $\ket{S0}$. In the limit of vanishing interdot tunneling, $t\ll\Gamma$, the nonlocal 
	singlet occupation and, therewith, these deviations cease. At finite magnetic field the lowest eigenenergies of the singlet 
	($s=0$), the doublet ($s=1/2$), and  the triplet sector ($s=1$) are shifted by $-\abs{sB}$. Hence, for sufficiently large magnetic 
	field, the system will be in a  current suppressing triplet ground state. For intermediate magnetic fields, the Josephson current can feature
	singlet--doublet (dotted line), triplet--doublet (dot-dashed line), and singlet--triplet (solid line) quantum phase transitions, whereby the doublet 
	ground state reverses the flow of Cooper pairs, $J<0$.
	
	In the following, we study the effect of a finite resonator coupling $\lambda$ and focus on the configuration exhibiting 
	a triplet--doublet transition (dot-dashed line in Fig.~\ref{fig.:phaseTransition}). To this end let us revisit 
	Fig.~\ref{fig.:scheme}(b) showing the phase diagram of the Josephson current as a function of $\lambda$ and $\phi$. 
	One recognizes immediately that it is subdivided into three different regions with sharp transitions at zero temperature.
	These regions correspond to a triplet (white region), a doublet (blue region), and a singlet ground state (red region) of the
	QD system. For large enough $\lambda$ both, the triplet and the doublet ground state, change to a singlet ground state
	featuring a positive Josephson current $J$. Eventually, this current abates for large resonator coupling, $\lambda\gg\hbar\omega$. 
	Intriguingly, one also observes the occurrence of a \textit{triple point}, where all three regions have a common point of intersection.
	It may qualify for high precision measurements of the resonator coupling strength $\lambda$.
	
	\begin{figure}[t]
		\includegraphics{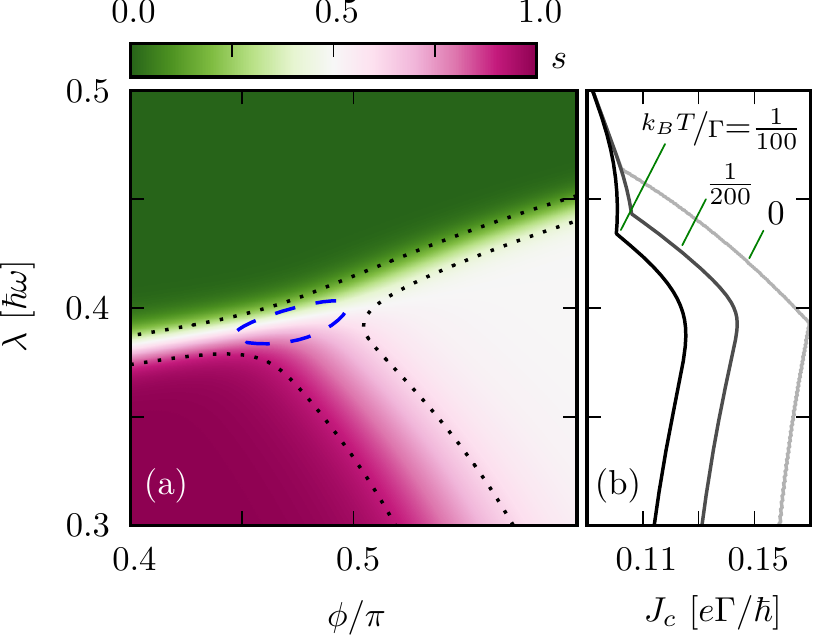}		%
		\caption{\label{fig.:finiteT}
		(a) Total spin quantum number $s$ as a function of the Josephson phase $\phi$ and the coupling $\lambda$ for
		$k_BT=\Gamma/200$ and all other parameters as in Fig.~\ref{fig.:scheme}(b). The dashed
		and dotted lines indicate the full width at half maximum of the overlaps $\Pi_{\rm tripel}$ and $\Pi_{\rm trans}$ 
		defined in Sec.~\ref{sec.:thermalBroadening}, characterizing the broadening of the triple point and the phase transitions, respectively.  
		(b) Critical current $J_c$ in dependence of $\lambda$ for different temperatures $T$.
		}
	\end{figure}
	We will demonstrate hereafter, that such triple point also manifests in the critical current $J_c=\max_\phi J(\phi)$, which is 
	experimentally easier accessible. Hence, we depict in Fig.~\ref{fig.:JCritical}(a) the critical current in dependence of the resonator 
	coupling $\lambda$ and for different magnetic fields $B$. For a sufficiently large coupling $\lambda$, the critical current $J_c$ becomes independent 
	of the magnetic field, since the DQD assumes then an unmagnetic singlet ground state, as discussed above. For small $\lambda$, however, 
	an increasing magnetic field $B>B_c$ with $B_c\approx0.75\Gamma$ reduces the critical current indicating, thus, the presence of a
	triplet ground state. In particular, the critical current in $\lambda$ shows up to two characteristic kinks, which for $B=0.8\Gamma$ (purple line) 
	are roughly located about $\lambda\approx0.24\Gamma$ and  $\lambda\approx0.41\Gamma$. For magnetic fields below $B_c$, however, the left kink disappears.
	To better understand the nature of these kinks, we inspect for $B=0.8\Gamma$ in panel (b) the corresponding lowest eigenvalues of each sector.
	One sees, indeed, that the left kink is associated to the triple point and that the right kink corresponds to a transition from a mixed singlet--doublet 
	ground state to a pure singlet ground state.
	In the inset of panel (a), we show the coupling strength $\lambda_{\textrm{tri}}$ at which, for a given magnetic field $B$, a triple point emerges.
	The required coupling strength $\lambda_{\textrm{tri}}$ increases above $B_c$ with the magnetic field $B$ (black solid line). Below $B_c$, however, 
	no triple point is possible. The dashed gray line indicates the estimated condition for the triple point given in the Appendix. It agrees well
	with the full calculation, but slightly underestimates the critical Zeeman splitting $B_c$.
	
	\section{\label{sec.:thermalBroadening}Thermal broadening of phase transitions}
	While the phase transitions at zero temperature are driven by quantum fluctuations, thermal fluctuations additionally contribute at finite temperature. 
	Their main effect is to smooth out these phase transitions, which is already observable at typical base temperatures of $T\approx 15$--$50\ \mathrm{mK}$
	being small against $\Gamma/k_B$ (ranging from $0.1$--$1\ \mathrm{meV/k_B}\approx 1$--$10\ \mathrm{K}$).~\cite{JorgensenNanoL2007a,DelagrangePRB2016a,EstradaSaldanaPRL2018a}
	
	For instance, this can be appreciated in figure~\ref{fig.:finiteT}(a) showing the total spin quantum number 
	$s$ obtained from the thermal average $\ev{\bm{S}}_{\textrm{POL}}^2=\hbar^2 s (s+1)$ over the electronic subsystem.~\footnote{
	The thermal average $\ev{\ldots}_{\textrm{POL}}\equiv \tr\big(\ldots\rho_{\textrm{POL}} \big)$ over the electronic subsystem is given
	with respect to the thermal state $\rho_{\textrm{POL}}\propto\exp\big(-H_{\textrm{POL}}/k_BT\big)$
	}  Here, $\bm{S}=(\hbar/2)\sum_{\alpha ss'} d_{\alpha s}^\dag \bm{\sigma}_{ss'} d_{\alpha s'}$ denotes the spin operator with $\bm{\sigma}$ 
	being the vector of Pauli matrices. The dotted black lines in Fig.~\ref{fig.:finiteT}(a)  highlight the thermal broadening, indicating the full width at half maximum of  
	$\Pi_{\rm trans}\equiv P_S P_T+P_S P_D+P_D P_T$ with $P_k=\tr_k\rho_{\textrm{POL}}$ being the cumulative populations of each sector $k=S,T,D$.
	In particular, the triplet--doublet transition is stronger effected by thermal excitations than transitions between magnetic and unmagnetic states. Similarly, 
	the overlap $\Pi_{\rm tripel}\equiv P_S P_T P_D$ gives a notion of the thermal broadening of the triple point---its full width at half maximum is indicated 
	by the dashed blue contour line. For increasing temperature, $T\ll\Gamma/k_B$, the phase-space volume of $\Pi_{\rm tripel}$ grows roughly quadratically.
	The critical current $J_c$ (panel b) mainly reduces with growing temperature but preserves its peak about $\lambda\approx0.4\hbar\omega$ 
	corresponding to the broadened triple point. 
	
	\begin{table*}[t]
		\begin{ruledtabular}
			\begin{tabular}{ll}
				Hamiltonian & basis \\
				\hline
				$H^{\textrm{singlet}}= 
				\begin{pmatrix}
					2(\epsilon-\epsilon_0) & 2(\epsilon-\epsilon_0) & 0 & 0 & 0\\
					2(\epsilon-\epsilon_0) & 2(\epsilon-\epsilon_0) & 0 & -i\bar\Gamma\sin\frac{\phi}{2} & \bar\Gamma\cos\frac{\phi}{2}\\
					0 & 0 & 2\epsilon & 0 & -t\\
					0 & i\bar\Gamma\sin\frac{\phi}{2} &  0 & 2(\epsilon-\epsilon_0 -\bar U) & 0\\
					0 &  \bar\Gamma\cos\frac{\phi}{2} & -t & 0 & 2(\epsilon-\epsilon_0 -\bar U)
				\end{pmatrix}$ 
				& 
				$ \!\!\begin{array}{l}\big\{ 
					\dfrac{\ket{dd}-\ket{0}}{\sqrt{2}}, \dfrac{\ket{dd}+\ket{0}}{\sqrt{2}},\\\;\;
					\ket{S0}, \ket{S-}, \ket{S+}
				\big\}\end{array}$
				\\[1ex]
				$H^{\textrm{triplet}} = 
				\diag(2\epsilon+B,\; 2\epsilon,\; 2\epsilon-B)$ & $\big\{ \ket{T\up}, \ket{T0}, \ket{T\down}\big\}$ 
				\\[1ex]
				$H^{\textrm{doublet}} = 
				\begin{pmatrix}
					\frac{B}{2}\id_2\otimes\sigma_z +\bar\epsilon\;\id_4 +\frac{t}{2}\sigma_x\otimes\id_2 
					& \frac{\bar\Gamma}{2}\exp\big[-i\frac{\phi}{2}\sigma_z\otimes\id_2]\\
					\frac{\bar\Gamma}{2}\exp\big[i\frac{\phi}{2}\sigma_z\otimes\id_2] 
					& \frac{B}{2}\id_2\otimes\sigma_z +\big[\bar\epsilon+2(\epsilon-\epsilon_0)\big]\id_4 -\frac{t}{2}\sigma_x\otimes\id_2
				\end{pmatrix}$ 
				&	 $ \!\!\begin{array}{l}\big\{ 
					\ket{R\up},\ket{R\down}, \ket{L\up},\ket{L\down}, \\\;\;
					\ket{tR\up},\ket{tR\down}, \ket{tL\up},\ket{tL\down} 
				\big\}\end{array}$
			\end{tabular}
		\end{ruledtabular}
		\caption{Decomposition of the effective Hamiltonian  $H_{\textrm{POL}}|_{k_BT=0}=H^{\textrm{singlet}}\oplus H^{\textrm{triplet}} \oplus H^{\textrm{doublet}}$
		at zero temperature into sectors of  different total spin. Here, $\sigma_x$, $\sigma_y$, $\sigma_z$ denote the Pauli matrices, $\id_n$ is the identity matrix 
		in the dimensions $n\times n$, and $\epsilon_0=-U_C/2$. 
		\label{tab.:sectors}
		}
	\end{table*}   
	
	\section{Conclusions}
	The Josephson transport through a carbon nanotube DQD circuit has been investigated. We have shown that the total spin of the DQD system characterizes 
	at zero temperature its ground state and, therewith, the Josephson current through the circuit. For the latter, we have derived an analytical estimate. While singlet 
	and doublet ground states entail a current phase relation of opposite sign, a triplet ground state rather suppresses the Cooper-pair tunneling. We have demonstrated 
	that a finite coupling to the resonator can, indeed, induce quantum phase transitions between these ground states and lift the triplet blockade. A large resonator coupling 
	eventually drives the system into a singlet ground state. In particular, we found that the resonator can induce a triple point in the Josephson current and analyzed 
	under which conditions this happens. Further, we have seen that such triple point also leaves it's footprint in the critical current. For experimentally relevant temperatures, 
	thermal fluctuations occur in addition to the quantum ones. Overall, a finite temperature reduces the supercurrent and washes out the  triple point and, hence, the 
	quantum phase transitions. The characteristic transport features, however, still remain. 
	
	The proposed hybrid device is an ideal platform for accessing the ultimate limit of a mechanical resonator that is coupled to Cooper pairs traversing a 
	Josephson junction. It not only constitutes a quantum hybrid device based on the fundamental coupling between  mechanical and electronic degrees of freedom but also paves the way for ultra-sensitive displacement detectors. A future perspective might be to go beyond the approximation of a thermal resonator state and investigate time-dependent dynamics that is controlled by quantum fluctuations.

	\begin{acknowledgments}
		We thank Daniel Reger for contribution to the calculations and Raffael Klees for helpful discussions. R.H. acknowledges financial support from the Carl-Zeiss-Stiftung. 
		W. B. was supported by the Deutsche Forschungsgemeinschaft (DFG, German Research Foundation) Project-ID 32152442 - SFB 767 and Project-ID 425217212 - SFB 1432. 
	\end{acknowledgments}
	
	\appendix
	\section{Josephson current for weakly coupled superconductors} 
	In this section, we provide an estimate of the Josephson current in the limit of weakly coupled superconductors. 
	Since the effective Hamiltonian $H_{\textrm{POL}}$ of the main text is block-diagonal in the sectors of different total spin,   
	see table~\ref{tab.:sectors}, each sector can be diagonalized independently. In the following, we calculate the 
	lowest eigenvalue of each sector in order to determine the groundstate energy of the system. Therefrom, we
	derive the desired estimate of the Josephson current.
	In the following, we restrict ourselves to the particle-hole symmetric point, $\epsilon=\epsilon_0$, and see from table~\ref{tab.:sectors} that
	\begin{align}
	E_T &=2\epsilon_0-\abs{B}\label{eq.:ET}
	\end{align}
	is the lowest eigenvalue of the triplet sector.
	To find the lowest eigenvalue of the doublet sector, we rewrite the corresponding Hamiltonian as
	\begin{align}
	&H^{\textrm{doublet}}\big|_{\epsilon=\epsilon0} = 
	\bar\epsilon\;\id_8 + \id_4\otimes\Big[\frac{B}{2}\sigma_z\Big]\\
	&+\Big[
		\frac{t}{2}\sigma_z\otimes\sigma_x
		+\frac{\bar\Gamma\cos\frac{\phi}{2}}{2}\sigma_x\otimes\id_2
		+ \frac{\bar\Gamma\sin\frac{\phi}{2}}{2}\sigma_y\otimes\sigma_z 
	\Big]\otimes\id_2 \nonumber\label{eq.:HDPH}
	\end{align}
	with $\sigma_k$ Pauli matrices.
	One immediately recognizes, that the first term just shifts the eigenspectrum by $\bar\epsilon$.
	The latter two terms are of the form of a Kronecker sum, 
	$\id_{\textrm{dim} Y}\otimes X + Y\otimes\id_{\textrm{dim} X}$, with its 
	eigenvalues composed of all pairs 
	of the eigenvalues of $X$ and $Y$~\cite{Laub2004a,Bjoerck2015a}.
	While the term corresponding to $X$ is already diagonal with the eigenvalues $\pm B/2$, 
	the term corresponding to $Y$ features a  biquadratic characteristic equation with the four 
	eigenvalues $\pm\frac{1}{2}\sqrt{{\bar\Gamma}^2+t^2\pm 2\bar\Gamma t\sin(\phi/2)}$;
	both plus-minus signs are independent of each other. By collecting all the mentioned
	contributions, we find
	\begin{equation}
		\begin{split}
			E_D 
			&= \bar\epsilon - \frac{\abs{B}}{2} -\frac{1}{2}\sqrt{{\bar\Gamma}^2+t^2 +2\bar\Gamma t\abs{\sin(\phi/2)}}
		\end{split} 
		\label{eq.:ED} 
	\end{equation}
	to be the lowest eigenvalue of the doublet sector. The modulus in the last term takes into account the
	change of sign of the sine function over a period.
	
	Finally, we estimate the lowest eigenvalue of the singlet sector. Firstly, one
	observes that $\frac{1}{\sqrt{2}}(\ket{dd}-\ket{0})$ becomes a zero-eigenstate at the particle-hole symmetric point.
	So, the remaining spectrum can be found from the reduced Hamiltonian
	\begin{align}
	H_{\bar\Gamma} &\equiv
			\begin{pmatrix}
	0&0&-i\bar\Gamma\sin\frac{\phi}{2}&\bar\Gamma\cos\frac{\phi}{2}\\
	0&2\epsilon_0&0&-t\\
	i\bar\Gamma\sin\frac{\phi}{2}&0&-2\bar U&0\\
	\bar\Gamma\cos\frac{\phi}{2}&-t&0&-2\bar U
			\end{pmatrix}.
	\end{align}
	In absence of the superconductors, $\bar\Gamma=0$, its unperturbed eigenvalues read
	$E_0^0 = 0$, $E_1^0 = 2\bar U$, $E_2=\epsilon_0-\bar U+\sqrt{(\epsilon_0+\bar U)^2+t^2}$, and
	\begin{align}
	E_{S}^0 &=\epsilon_0-\bar U-\sqrt{(\epsilon_0+\bar U)^2+t^2}
	\label{eq.:Epm},
	\end{align}
	satisfying $E_2^0,E_0^0>E_1^0>E_S^0$ for $t,\bar U>0$ and $\epsilon_0<0$.
	Hereafter, we will assume that $t^2\neq-4\epsilon_0\bar U$ for which $E_2^0$ and $E_0^0$ become non-degenerate.
	In the limit of weakly coupled superconductors, one can formally expand the lowest eigenvalue of the 
	singlet sector $E_S\equiv E_S(\bar\Gamma)$ to the second order in $(\bar\Gamma/E_S^0)$,
	\begin{align}
	E_S(\bar\Gamma) \approx E_S(0)+ \bar\Gamma E'_S(0)  + \frac{{\bar\Gamma}^2}{2} E''_S(0),
	\label{eq.:ESTaylor}
	\end{align}
	where $E_S(0)\equiv E_S^0$.
	The Taylor coefficients can be related to the characteristic polynomial $P(\bar\Gamma,\lambda) \equiv\det\big(H_{\bar\Gamma} -\lambda\id_4\big)$ which,
	in particular, vanishes at the eigenvalue $\lambda=E_S(\bar\Gamma)$. Since also its derivatives have to
	vanish, one finds from the relation $0=\partial_{\bar\Gamma}P\bm{(}\bar\Gamma,E_S(\bar\Gamma)\bm{)}$ that $E'_S(0)$ is zero.
	Similarly, one finds from the second derivative in $\bar\Gamma$ of the characteristic equation, 
	$0=\partial_{\bar\Gamma}^2 P\bm{(}\bar\Gamma,E_S(\bar\Gamma)\bm{)}$ and under consideration of $E'_S(0)=0$ the relation
	\begin{align}
	E''_S(0) &=-\frac{P^{(2,0)}(0,E_S^0)}{P^{(0,1)}(0,E_S^0)}
	= \frac{
		(1+\cos\phi)(2\epsilon_0-E_S^0)
	}{
		2E_S^0(E_S^0 -\epsilon_0+\bar U)
	}.
	\end{align}
	In the last step, we evaluated the derivatives of the characteristic polynomial,
	\begin{align}
	P(\bar\Gamma,\lambda) ={ }&\frac{{\bar\Gamma}^2}{2}(1+\cos\phi)(2\epsilon_0-\lambda)(2\bar U+\lambda)\nonumber\\ 
	+&\frac{r(\lambda)}{2}\big[
		2\lambda(2\bar U+\lambda) + {\bar\Gamma}^2(\cos\phi-1)
	\big]
	\label{eq.:charPol},
	\end{align}
	exploiting that the prefactor $r(\lambda)\equiv(\lambda-2\epsilon_0)(\lambda+2\bar U)-t^2$ vanishes at $\lambda=E_S^0$.
	Thus, equation~\eqref{eq.:ESTaylor} yields the approximation
	\begin{align}
	E_S &\approx E_S^0 + \frac{
		\bar\Gamma^2\big(1+\cos\phi\big)\big(E_S^0-2\epsilon_0\big)
	}{
		4E_S^0 \big(E_S^0 -\epsilon_0+ \bar U \big)
	} \label{eq.:ESApprox}
	\end{align}
	for the lowest eigenvalue of the singlet sector.
	
	Finally, one obtains from the derivative of the groundstate energy $E_{\textrm{GS}} = \min\{E_S,E_D, E_T\}$ with respect
	to the Josephson phase $\phi$ the estimate of the Josephson current Eq.~\eqref{eq.:JEst} given in the main text.
	Moreover, from $E_T=E_D$ and $E_T=E_S$ one finds the equations
	\begin{align}
	\sin^2\frac{\phi}{2} &=  \Bigg[\frac{\bar\Gamma^2+t^2 -(2\epsilon_0+\bar U -\abs{B})^2}{2\bar\Gamma t}\Bigg]^2, \\
	\cos^2\frac{\phi}{2} &= \frac{2E_S^0(\epsilon_0-\bar U-E_S^0)(E_S^0-2\epsilon_0+\abs{B})}{\bar\Gamma^2(E_S^0-2\epsilon_0)},
	\end{align}
	which, added up, $\sin^2(\phi/2) +\cos^2(\phi/2)=1$, yield a condition for the triple point.

\end{document}